\pdfoutput=1

\documentclass[journal]{IEEEtran}
\usepackage{graphicx}
\usepackage{epstopdf}
\usepackage{epsfig,subfigure}
\usepackage{amssymb,amsmath}
\usepackage{enumerate}
\usepackage{multirow}
\usepackage{array}
\usepackage{cite}
\usepackage{makecell}
\usepackage{mathrsfs}
\usepackage{amsmath}

\usepackage[noend]{algpseudocode}
\usepackage{algorithmicx,algorithm}
\usepackage{soul}
\soulregister\cite7 
\soulregister\ref7 
\usepackage{color}

\begin{document}

\title{
Failure-Resilient\! Distributed\! Inference\! with\! Model Compression over Heterogeneous Edge Devices

\author{Li Wang, \emph{Senior Member}, \emph{IEEE}, Liang Li, \emph{Member}, \emph{IEEE},  Lianming Xu, Xian Peng, and Aiguo Fei 
\thanks{Li Wang, Xian Peng, and Aiguo Fei are with the School of Computer Science (National Pilot Software Engineering School), Beijing University of Posts and Telecommunications (BUPT), China (Email: liwang@bupt.edu.cn; pixie@bupt.edu.cn; aiguofei@bupt.edu.cn). Liang Li is with the Frontier Research Center, Peng Cheng Laboratory, China (Email: lil03@pcl.ac.cn). Lianming Xu is with the School of Electronic Engineering, Beijing University of Posts and Telecommunications (BUPT), China (Email: xulianming@bupt.edu.cn). (\textit{Corresponding author: Liang Li}) } 
\thanks{This work was supported by National Natural Science Foundation of China under grants 62201071 and 62171054. }
}
}

\maketitle

\begin{abstract}
The distributed inference paradigm enables the computation workload to be distributed across multiple devices, facilitating the implementations of deep learning based intelligent services on extremely resource-constrained Internet of Things (IoT) scenarios. Yet it raises great challenges to perform complicated inference tasks relying on a cluster of IoT devices that are heterogeneous in their computing/communication capacity and prone to crash or timeout failures. In this paper, we present RoCoIn, a robust cooperative inference mechanism for locally distributed execution of deep neural network-based inference tasks over heterogeneous edge devices. It creates a set of independent and compact student models that are learned from a large model using knowledge distillation for distributed deployment. In particular, the devices are strategically grouped to redundantly deploy and execute the same student model such that the inference process is resilient to any local failures, while a joint knowledge partition and student model assignment scheme are designed to minimize the response latency of the distributed inference system in the presence of devices with diverse capacities. Extensive simulations are conducted to corroborate the superior performance of our RoCoIn for distributed inference compared to several baselines, and the results demonstrate its efficacy in timely inference and failure resiliency.
\end{abstract}
\begin{keywords}
Edge intelligence, Internet of Things, distributed deep learning inference, knowledge distillation. 
\end{keywords}

\IEEEpeerreviewmaketitle

\section{Introduction}
Exciting breakthroughs in deep learning (DL) and Internet of Things (IoT) have opened up new possibilities for pervasive intelligence at the network edge. This is achieved through the deployment of deep neural networks (DNNs) on various mobile edge devices, in response to the growing need for on-device intelligent services across a wide range of application domains, ranging from intelligent assistants (such as Google Now and Amazon Echo) in smart home to advanced video analytics in smart cities. Basically, the outstanding performance of DNNs in accurate human-centric content processing notoriously relies on increasingly complex parameterized models that are memory-hungry and computation-intensive, which poses significant challenges to implement DNNs on embedded IoT devices with limited resources (e.g., memory, central processing units (CPUs), battery, bandwidth). For instance, microcontrollers like Arm Cortex-M, commonly used in IoT applications such as Smart Healthcare and Keyword Spotting, typically have limited available memory, typically around 500KB. As a result, models like Residual Network-50 (ResNet-50) with 50 convolutional layers, which require over 95 megabytes of memory for storage and involve multiple floating-point multiplications for each image calculation, may not be feasible for deployment on such devices. Therefore, there is a fundamental need for alternative approaches that are more memory-efficient and computationally lightweight to enable fast and IoT-devices friendly DNN inference. 



To alleviate the computation burden of edge devices for DNN inference, some model compression techniques exist in the literature, wherein Knowledge Distillation (KD) stands out as an effective solution to produce a more compact model that can substitute for a complex model \cite{cheng2018model}. Specifically, KD-based methods train a more compact neural network (a.k.a student model) with far fewer layers/width to mimic the output of a larger network (a.k.a teacher model) we want to compress \cite{gou2021knowledge}. The basic idea behind KD is to distill the knowledge from the teacher model into the student model, using a distillation algorithm that is optimized for imitation performance. This process involves three essential components: knowledge, distillation algorithm, and teacher-student architecture, which have been improved by researchers from both theoretical and empirical perspectives. However, a lightweight student model that is compatible with extremely resource-constrained IoT devices may not have the necessary capacity to represent the teacher's knowledge, thereby suffering from severe accuracy loss. 


Several pioneering works propose to exploit available computation resources within a manageable range for distributed inference, instead of keeping all computation at the single local device, under the emerging edge intelligence paradigm \cite{hu2019dynamic,xue2020edgeld,mohammed2020distributed,zhang2019edge,schlegel2022privacy}. As such, by forming a collaborative DNN computing system, the inference workload is partitioned and distributed from the source device to a cluster of devices in proximity via local wireless connections such as WiFi. Jouhari et al. in \cite{jouhari2021distributed} divide the layers of the given DNN into multiple subsets, each of which is executed on a separate device, with intermediate feature maps transferred between the corresponding devices at runtime. Mao et al. in \cite{mao2017modnn} enables execution parallelism among multiple mobile devices by partitioning the neurons of each layer where the overlapped parts of the layer inputs need to be transferred among the devices during computation. To alleviate the prohibitively huge communication burden for intermediate exchanges, Bhardwaj et al. in \cite{bhardwaj2019NoNN} take this approach a step further and designed a new distributed inference paradigm called Network of Neural Networks (NoNN), which compresses a large pre-trained `teacher' deep network into a set of independent `student' networks via KD. These individual students can then be deployed on separate resource-constrained IoT devices to perform the distributed inference. In this way, only the outputs of the student networks require aggregation for inferring the final result, thereby reducing communication overhead to a large extent.

Despite the benefits of parallelized computational workloads and negligible accuracy loss, there are still a few issues that hinder the efficient execution of DNN inference tasks over massive resource-constrained edge devices. On the one hand, edge devices usually have heterogeneity in their computation capacity and communication condition, thereby disabling those distributed inference schemes that uniformly partition the knowledge of the teacher model and transfer it to the student models with the same structure. It is non-trivial to distribute the inference workload over heterogeneous devices for timely response with full utilization of edge resources while alleviating the bottleneck effect from the stragglers with insufficient capacities. On the other hand, since the cooperative mechanism parallelizes DNN inference in a distributed manner, the crash of any edge device or network timeout can result in system breakdown and invalidate the inference result. Such failures are unknown to the task requester a priori and hard to harness proactively for maintaining the inference performance. Therefore, it fundamentally calls for the development of more flexible and adaptable approaches to partitioning the knowledge of the teacher model and distributing it to the student models, as well as designing robust cooperative inference systems that can handle the failure of individual edge devices without compromising the responsiveness or accuracy of the overall system. Achieving these goals will be critical to realizing the full potential of edge intelligence and enabling efficient and scalable distributed inference on resource-constrained edge devices.

In this paper, we develop a failure-resilient model compression and distribution scheme, named RoCoIn, for cooperative deep learning model inference. Our scheme employs a similar parallel workflow as NoNN where a set of independent student models are distilled from the large model for distributed deployment. In particular, RoCoIn enables knowledge dissemination across heterogeneous devices and ensures the resilience of the cooperative inference system against local failures. With a focus on minimizing response latency, RoCoIn strategically groups devices to redundantly handle the same student model with a resilience guarantee, while incorporating a joint knowledge partition and model assignment method to accommodate devices' diverse capacities without sacrificing inference accuracy. As a result, devices can deploy individual student models with varying complexities for parallel computing, with only the outputs requiring aggregation to infer the final result. RoCoIn lays the groundwork for intermediary interaction-free cooperative inference across heterogeneous edge devices and sets the stage for further enhancements and integrations with other resource scheduling policies. 
Our salient contributions are listed as follows: 

\begin{itemize}
\item  We present RoCoIn, a robust cooperative inference mechanism to enable failure-resilient distributed execution of deep neural network-based inference tasks via local knowledge replication. It distills the knowledge of the deep model to a set of independent and compact models, while enabling edge devices' local deployments with knowledge replication.

\item  We propose a knowledge assignment algorithm that partitions the knowledge of the original deep model with importance balancing and designates the target locally-deployed models for minimizing the inference latency with the accuracy guarantee. Accommodating the edge devices' diverse capacities (i.e., the processors' computational performance, the memory budget, and the transmission quality), the algorithm integrates the similarity-aware device grouping and the normalized cut-based knowledge partitioning, where the Kuhn-Munkres method is further used to obtain the optimal device-knowledge-student matching.

\item We evaluate the performance of the RoCoIn with our knowledge assignment scheme via simulations, which verify the efficacy of our scheme with various data sources and system configurations. Several baselines are conducted to validate the superiority of RoCoIn in terms of timely inference and failure resiliency.

\end{itemize}

The remainder of the paper is organized as follows. Section II reviews related work on edge inference. Section III elaborates on the RoCoIn design and builds the system model. Section IV formulates the knowledge assignment problem and presents our algorithm. Section V gives the performance evaluation, and Section VI finally concludes the paper.

\section{Related Work}

\subsection{Model Compression}
To enable DNN training or inference relying on resource-limited IoT devices, one possible solution could be to use more optimized and compact deep learning architectures specifically designed for these devices, subjecting to precious memory and processing power. This motivates some model compression techniques in the literature to enhance training efficiency or inference efficiency~\cite{wang2022edcompress,wang2020context,li2020talk}. While they operate at different stages of the machine learning pipeline and the objectives may differ between the two scenarios, both aim to reduce the computational resources required by the model. As a result, many model compression techniques (such as pruning, quantization, knowledge distillation, etc.) can be adapted or extended to address both training and inference efficiency. The rationale behind is that deep models usually learn a large number of redundant or useless weights, which can be removed or compactly represented while sacrificing accuracy to a moderate extent. Specifically, pruning and quantization aim to reduce the number of weights and the number of bits required to represent weights in deep networks, respectively\cite{han2015compression}. KD-based methods aim to train a more compact neural network (a.k.a student model) with far fewer layers/width to mimic the output of a larger network (a.k.a teacher model) we want to compress \cite{hinton2015distillation}. Using an ensemble KD technique, Bharadhwaj et al. in \cite{bharadhwaj2022detecting} improved the performance of tiny vehicle detection models that enables edge device-based vehicle track and count for real-time traffic analytics. Xu et al. in \cite{xu2021kdnet} proposed a hybrid KD framework to compress a complex long short-term memory model for machine remaining useful life, which includes a generative adversarial network based knowledge distillation for disparate architecture knowledge transfer and a learning-during-teaching based knowledge distillation for identical architecture knowledge transfer. However, the ability of model compression to reduce resource overhead is limited for those IoT devices with extremely constrained capacity, and excessive compression will result in severe degradation of model performance in intelligent data understanding.


\subsection{Cooperative Edge Inference}
Some recent efforts have been devoted to mitigating the computational bottleneck at single edge device by coordinating multiple devices to jointly perform intelligent tasks via proper model/data partition and workload distribution, facilitated by controllable inter-device communication \cite{hu2019dynamic,xue2020edgeld,mohammed2020distributed,zhang2019edge,schlegel2022privacy,mao2017modnn,zhao2018deepthings}. Neurosurgeon \cite{kang2017neurosurgeon} first proposed to partition a DNN model and offload partial inference workload onto a powerful cloud server for follow-up execution, where the offloading efficiency highly depends on the transmission condition of unstable wide-area network connections. To fully utilize the decentralized resources of massive edge devices, Jouhari et al. in \cite{jouhari2021distributed} divided the layers of the given DNN into multiple subsets, each of which is executed on a separate device and the intermediate feature maps are transferred between the corresponding devices at runtime. Considering the complex model architecture with a directed acyclic graph (DAG) rather than a chain of layers, Hu et al. in \cite{hu2022distributed} presented EdgeFlow to enable distributed inference of general DAG structured DNN models. EdgeFlow partitions model layers into execution units and orchestrates the intermediate results flowing through these units to fulfill the complicated layer dependencies. Such model-parallelism paradigms are particularly beneficial when the model is too large to fit on a single device's memory or when certain model components require specialized hardware for efficient computation.

Instead of partitioning the model, Zeng et al. in \cite{zeng2020coedge} proposed to split the input data for matching the available resources of edge devices, which does not sacrifice model accuracy as it reserves input data and model parameters of the given DNN model. Data-parallelism schemes can be effective when the model size is manageable, and the input data can be easily partitioned into smaller batches. Yet they assume that edge devices are powerful in their memory to accommodate the entire DNN model, which may hinder its applicability. To alleviate the prohibitively huge communication burden for intermediates exchange while conserving memory of edge devices, Bhardwaj et al. in \cite{bhardwaj2019NoNN} designed a new distributed inference paradigm, named Network of Neural Networks (NoNN), that compresses a large pre-trained `teacher' deep network into a set of independent student networks via KD. These individual students can then be deployed on separate resource-constrained IoT devices to perform the distributed inference, where only the outputs of the student networks require aggregation for inferring the final result. However, NoNN uniformly distributes the learned knowledge into student models with identical structures, which are weak and vulnerable to adapt to the varying network conditions and capability among edge devices, especially in cases where some of the devices become stragglers due to crash or timeout issues.



\section{System Model and RoCoIn Workflow}


\begin{figure*}\centering 
  \centering
  \includegraphics*[width=.9\textwidth]{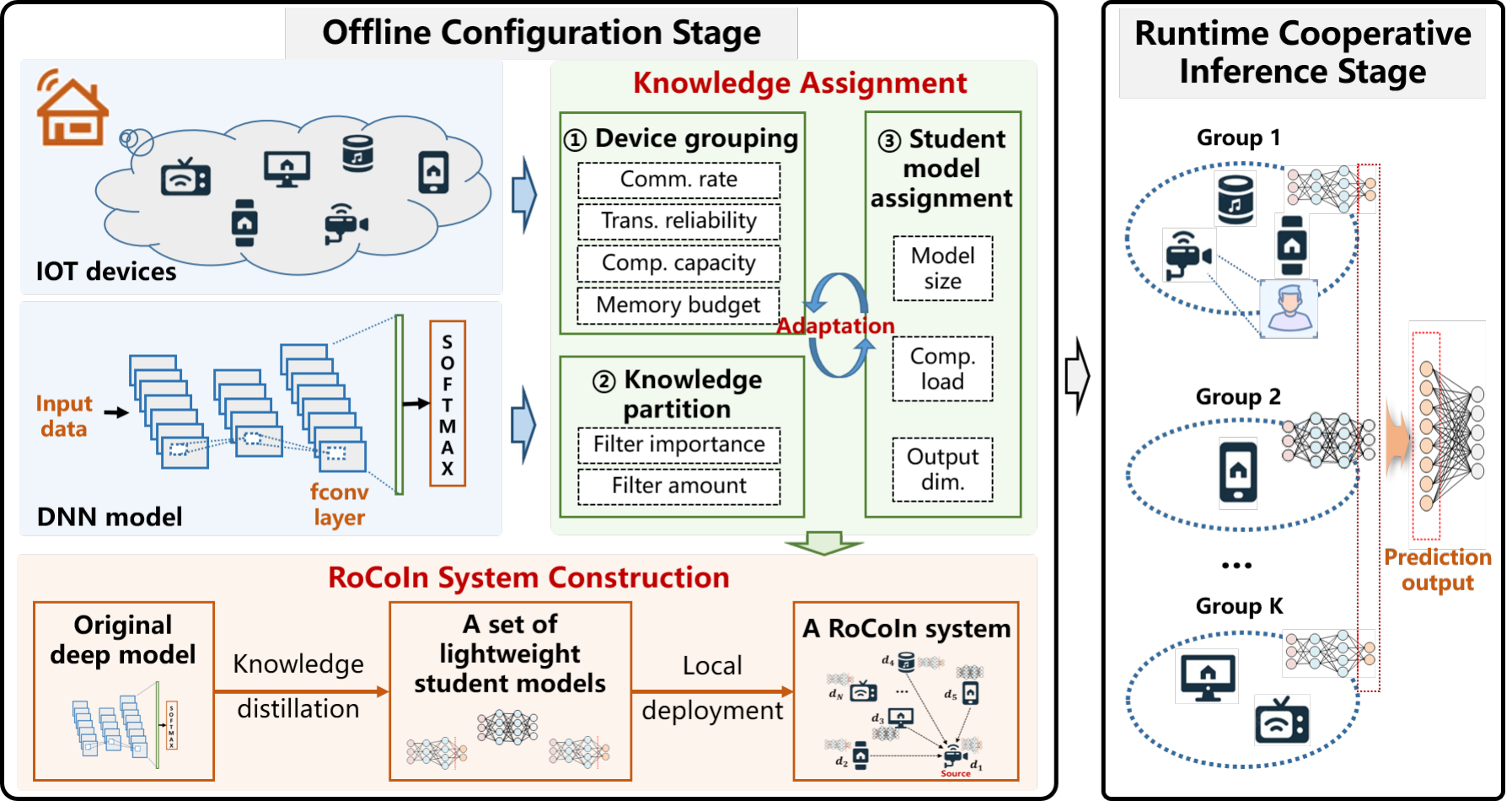}\\
  \caption{An illustration of RoCoIn mechanism.}\label{fig.system}
\end{figure*}

We describe our target distributed DNN inference system as follows with basic notations. We assume that there is a set of $N$ edge devices, denoted by $D=\{d_1,d_2,...,d_n,...,d_N\}$, distributed in a restricted area. In particular, for a certain type of inference task, we suppose that there is an edge device serving as a source, which broadcasts the raw data for cooperative computation and aggregates the parallelized outputs from the other devices for generating the final inference result. Taking the face recognition task in a smart home scenario as an example, the raw image is generally captured by an inspection camera that can serve as the source device and trigger the cooperative inference procedures for the task. We denote by $d_1$ the source device without loss of generality and suppose that $d_1$ can communicate with all other edge devices via wireless connection. Assume that the channel coefficient is Rayleigh distributed, i.e., $h_n\sim CN(0, \sqrt{\lambda})$, where the channel gain $g_n=|h_n|^2$ follows an exponential distribution. We use a tuple $(c_n^{core}, c_n^{mem}, r_n^{tran}, p_n^{out})$ to specify the resource profile of edge device $d_n$. Here, $c_n^{core}$ and $c_n^{mem}$ represent the $d_n$’s FLOP and memory budgets for inference tasks, respectively, reflecting the computing capability of $d_n$ in a coarse granularity. For a single device that only processes DNN workloads, $c_n^{mem}$ is the volume of memory excluding the space taken by the underlying system services, e.g., I/O services, compiler, etc. $r_n^{tran}$ denotes the the wireless transmission rate of the link $d_n \rightarrow d_1$, and $p_n^{out}$ represents its transmission outage probability.

We suppose that the teacher contains several convolutional layers and one or more full-connected (FC) layers for prediction. RoCoIn employs a similar parallel workflow as NoNN~\cite{bhardwaj2019NoNN}, where a collection of lightweight student models that focus only on a part of teacher’s knowledge are separately deployed on edge devices to perform the distributed inference. Thus, it parallelizes the execution on multiple devices at runtime. The key is to partition the knowledge of the teacher model, which can be achieved by clustering the filters in the teacher’s final convolution layer according to their activation patterns and using them to train individual student modules. The rationale behind is that features for various classes are learned at different filters in CNNs and the activation patterns reveal how teacher’s knowledge gets distributed at the final convolution layer. While NoNN create a filter activation network to characterize the distribution pattern of the teacher’s knowledge, RoCoIn proposes to optimize knowledge assignment over edge devices with appropriate student architecture selection against any unexpected failures during inference, adapting to heterogeneous edge resources and transmission conditions. We define $F=\{f_1,f_2,...,f_m,...,f_M\}$ as the set of filters belonging to the teacher’s final convolution layer. Let $\mathcal{P}=\{P_1,P_2,...,P_k,...,P_K\}$ be the set of filter partitions where $ P_k \subset F $ for any $k$. Assume that there are $J$ types of student architectures, denoted by $S=\{s_1,s_2,...,s_j,...,s_J\}$, with different computation load $R_j$(FLOPS) and memory requirement $Q_j$(bit), which can be selected to deployed on edge devices after learning certain knowledge from the teacher. 

Note here that, in practice, the edge devices that participated in the cooperative inference task are diverse in their types and functions, and the locally processed results may not be aggregated successfully due to uncertain system factors, e.g., edge device crash, unexpected channel conditions, concurrent computation tasks, etc. It may result in great damage to the performance of the distributed inference system since the source device is oblivious to the uncertainty a prior. To make the system failure-resilient, we propose to assign knowledge to the edge devices with replication, which allows multiple devices to undertake the same part of an inference task by considering the potential uncertainty in advance. Here, we use the set $\mathcal{G}=\{G_1,G_2,...,G_k,...,G_K\}$ to indicate the collection of edge device groups where $G_k \subset D$ for any $k$, and there will be a one-to-one matching between any device group and filter cluster. The main notations used throughout the paper are listed in Table \ref{notation}.

Fig.~\ref{fig.system} shows the workflow of our RoCoIn, which consists of the offline setup phase and the runtime execution phase. In the offline setup phase, RoCoIn records the execution profiles of each device and creates a cooperative inference plan that determines the knowledge partitions and their student model assignment using the knowledge assignment algorithm. Separate students are then trained to mimic parts of teacher’s knowledge, which are deployed on corresponding edge devices according to the cooperation plan to enable parallel execution. The runtime execution phase starts when the source device receives DNN inference query. As response, the source device establishes connections with the cooperative edge devices according to the cooperation strategy and distributes the input data, e.g., image, to them. All the participating devices feed the data into their local student model in parallel and generate a portion of the final convolution layer's output. These portions are aggregated by the source device and merged by a Fully Connected (FC) layer to yield the final prediction in response to the query. As the teacher’s knowledge is redundantly assigned to the students for resisting on any failures, the source device can initiate the FC-layer execution when receiving a necessary number of disjoint portions, rather than waiting for the feedback from all of the devices. Notice that our individual student models that are well selected adhere to the heterogeneous memory-and FLOP-constraints of the edge devices, and do not communicate until the final fully connected layer. Consequently, RoCoIn incurs significantly lower memory, computations, and communication, while improving the robustness by injecting redundancy in teacher’s knowledge assignment.

We note that the performance of our RoCoIn system strongly relying on appropriate knowledge assignment, wherein the following questions require to be answered: i) What replication rule should be used to determine the edge devices that act as backups for each other? 2) How to partition the teacher’s knowledge into disjoint portions catering to the intrinsic characteristics of the teacher model? 3) Which student architectures should be selected for the edge devices tailored to their diverse resource capacity, while fully learning corresponding knowledge partition? Notice that the three-fold strategies are closely intertwined, and thus there is a great demand for joint optimization to make full use of edge resources while drawing sufficient knowledge from the teacher, which will be elaborated in the subsequent section.

\renewcommand\arraystretch{1.3}
\begin{table*}[t]
  \small
  \centering
  \caption{SUMMARY OF NOTATIONS.}\label{notation}
  \begin{tabular}{|c|l||c|l|}
  \hline
  $D=\{d_n\}_{n=1}^N$ & Set of edge devices & $F=\{f_m\}_{m=1}^M$ & Set of filters in the final convolution layer\\ \hline 
  $\mathcal{P}=\{P_k\}_{k=1}^K$ & Set of filter patitions & $\mathcal{G}=\{G_k\}_{k=1}^K$ & Set of device groups \\ \hline  
  $c_n^{core}$ & FLOP budget of device $d_n$ & $c_n^{mem}$ & Memory budget of device $d_n$ \\ \hline
  $r_n^{tran}$ & Wireless transmission rate of device $d_n$ & $p_n^{out}$ & Transmission outage probability of device $d_n$\\ \hline
  $S=\{s_j\}_{j=1}^J$ & Set of available student models & $\alpha_{kj}$ & Binary variable, student assignment indicator\\ \hline
  $R_j$ & Computation load of student model $s_j$ & $Q_j$ &  Output size of student model $s_j$\\ \hline
  $p^{th}$ & Transmission failure probability threshold & $d^{th}$ & Device similarity threshold\\ \hline
  $\mathcal{F}$ & Filter activation pattern graph & $E=\{e_{mm'}\}_{\forall m,m'}$ & Edge set of graph $\mathcal{F}$\\ \hline
  $A=[A_{mm'}]_{\forall m,m'}$ & Weight matrix of graph $\mathcal{F}$ & $a_m$ & Average activity of filter $f_m$\\ \hline
  $z_m$ & Degree of node $f_m$ & $Z$ & Degree matrix of graph $\mathcal{F}$\\ \hline
  $W(P_k,P_{k'})$ & Cut weight of $P_k$ and $P_{k'}$ & $w(G_k,P_{k'})$ & Assignment weight of $G_k$ and $P_{k'}$\\ \hline
  $L$ &  Laplacian matrix of graph $\mathcal{F}$ & $H=[h_{mk}]_{\forall m,k}$ & Indicator matrix for filter partitioning\\ \hline
\end{tabular}
\end{table*}


\section{Knowledge Assignment Scheme Design}

In this section, we begin with the formulation of the knowledge assignment problem, followed by the elaboration on the design of the knowledge assignment scheme.

\subsection{Problem Formulation}

We first introduce a binary variable $\alpha_{kj}$ where $\alpha_{kj}=1$ if using student architecture $s_j$ to learn the knowledge regarding the filter partition $P_k$; Otherwise, $\alpha_{kj}=0$. On optimizing the knowledge assignment strategy, we aim to minimize the inference completion delay, in spite of some failures in the local output aggregation that may compromise the inference performance. Particularly, it requires us to make joint decisions on user grouping $\mathcal{G}=\{G_1,G_2,...,G_K\}$, filter partition $\mathcal{P}=\{P_1,P_2,...,P_K\}$, and student assignment $\alpha_{kj}$ under the constraints of heterogeneous edge resources. Towards this goal, we establish the knowledge assignment problem for our RoCoIn as:

\begin{subequations}  \label{IP}
\begin{align}
& \min_{\substack{K, \mathcal{G}=\{G_1,G_2,...,G_K\} \\ \alpha_{kj},\mathcal{P}=\{P_1,P_2,...,P_K\}}} \max \limits_{k} \min \limits_{n:d_n\in G_k} \sum \limits_{j} \alpha_{kj} (\frac{C_j^{flops}}{c_n^{core}} + \frac{Q_j}{r_n^{tran}}) \label{obj}  \\
        & s.t.\ \ \bigcup_k G_k = D, \label{group} \\
& \quad \quad \bigcup_k P_k = F, \label{filter} \\
& \quad  \quad G_k \bigcap G_{k'} = \emptyset, \forall k \neq k', k,k'\in\{1,...,K\}, \label{group-inter} \\
& \quad \quad P_k \bigcap P_{k'} = \emptyset, \forall k \neq k', k,k'\in\{1,...,K\}, \label{filter-inter} \\
& \quad \quad \prod \limits_{n:d_n\in G_k}p_n^{out} \leq p^{th}, \forall k\in\{1,...,K\}, \label{succ} \\
& \quad \quad \sum \limits_{j} \alpha_{kj} C_j^{para} \leq\! \min \limits_{n:d_n\in G_k}\!c_n^{mem},\!\forall k\in\{1,...,K\},\label{mem-cons}\\
& \quad \quad Loss(\theta_S|\theta_T)\leq \varepsilon^{th}.\label{error}
\end{align}
\end{subequations}

Here, the objective function in (\ref{obj}) represents the inference completion delay, which is blocked by the slowest group of devices that return the local output. Here, $C_j^{flops}/c_n^{core}$ calculates the execution delay for performing student model $s_j$ at edge device $d_n$, while $Q_j/r_n^{out}$ is the time consumed to transmit its output over wireless channels. The constraints in (\ref{group}-\ref{filter}) enforce that all the devices and the filters are partitioned for distributed inference. Constraints in (\ref{group}-\ref{filter}) further ensure that an edge device, as well as a filter, can be assigned to no more than one group/partition. (\ref{succ}) guarantees that the cumulative transmission failure probability of the devices in the same group should not exceed a threshold $p^{th}$, so that the portion of the output corresponding to each device group can be returned for aggregation with high-reliability guarantee. Constraints in (\ref{mem-cons}) enforce that the required memory budget to run student model on edge devices should not exceed their diverse capacity. Further, (\ref{error}) ensures that the student architecture selected for the edge devices can fully learning the teacher's knowledge without sacrificing accuracy, i.e., the loss attained by the student cluster is lower than the threshold $\epsilon^{th}$.

\subsection{Algorithm Design}
Considering the intertwined relation among variables, we decouple the original knowledge assignment problem in (\ref{IP}) and integrate three functional modules into the algorithm, i.e., device grouping, knowledge partition, and student assignment. The three modules run sequentially and determine decisions on device grouping $\mathcal{G}=\{G_1,G_2,...,G_K\}$, filter partition $\mathcal{P}=\{P_1,P_2,...,P_K\}$, and student assignment $\alpha_{kj}$ under the constraints of heterogeneous edge resources, respectively. To elaborate, the algorithm first groups the devices via modified follow-the-leader procedures based on a well-defined similarity distance. This process ensures that edge devices with similar computational capacities and satisfactory transmission reliability are clustered together to serve as replicas of each other. With the determined number of device groups, the filters from the teacher's final convolution layer are clustered into knowledge partitions of corresponding quantity. This is facilitated by constructing a weighted filter graph and optimizing through Normalized Cut. After that, the algorithm targets at achieve optimal matching among device groups, knowledge partitions, and student models, thereby minimizing inference delay while preserving accuracy. To reduce computational complexity, we simplify the three-dimensional matching into a bi-partite matching problem and integrate the KM algorithm to find its optimum. In the following, we detail the three modules in our knowledge assignment scheme. 

1) \textit{Device grouping}: We first define the capacity similarity of any two edge devices using Euclid distance, which is calculated by 

\begin{equation}
    sim(d_n,d_{n'}) = \sqrt{(c_n^{mem}-c_{n'}^{mem})^2+(c_n^{core}-c_{n'}^{core})^2},
\end{equation}

We use a modified follow-the-leader method to group the edge devices with approximately equal computational capacity, subjecting to the constraint in transmission reliability. The procedure does not require initialization of the number of device groups, and uses an iterative process to compute the cluster centroids. It starts by randomly setting a device as the centroid of the group $G_1$, which is denoted by $\overline{G_1}$. Then we calculate its capacity similarity $sim(\overline{G_1}, d_n)$ with the other devices $d_n\in D/G_1 $. The group $G_1$ involves the devices satisfying both $sim(\overline{G_1}, d_n)\leq d^{th}$ and $ \prod_{n:d_n\in G_k} p_n^{out} > p^{th}$ one by one, and recompute the centroid repeatedly. The devices that do not meet the conditions in terms of all the existing groups will be regarded as a new group, and all the groups continue to run the same procedures to involves the rest unassigned devices. The process is controlled by the distance threshold $d^{th}$, which is chosen through trial and error. 

2) \textit{Knowledge partition}: We distribute knowledge from teacher's final convolution layer to individual students for enabling parallel inference. We suppose that the teacher contains several convolutional layers and one or more FC layers for prediction. When passing an image from the validation set through the teacher network, each filter in the final convolution layer has a certain feature map. Inspired by \cite{bhardwaj2019NoNN}, we use the average activity metric as a measure of importance of a filter for a given class of images, which is defined as the averaged value of the corresponding output channel of teacher's final convolution layer. Basically, the higher the average activity of a filter, the more important it is for the classification for some classes of images. Let $a_m$ denote the average activity of a filter $m$ for a given image in the validation set. Then a weighted graph $\mathcal{F} = (F, E, A)$ of filter activation patterns can be built with filters $f_m\in F, \forall m$ as nodes, where each two nodes $(f_m, f_{m'}), m\neq m', f_m, f_{m'}\in F$ are connected by an edge $e_{mm'}\in E$ with $A_{mm'}=\sum_{val}a_m a_{m'}|a_m-a_{m'}|$ as the weight. 

On this basis, the partition of the filters for teacher's knowledge distribution can be regarded as a $K$-cut problem of the weighted graph
$\mathcal{F} = (F, E, A)$, which require us to split the graph into $K$ sub-graphs. For any filter partition $P_k$ corresponding to a sub-graph, we denote by ${vol(P_k)}$ the size of $P_k$. It is defined as ${vol(P)}=\sum_{f_m \in P} z_m$ where $z_m=\sum_{f_{m'}\in F}A_{mm'}$ is the degree of of the node $f_m$. We further define the weight of cut for any two disjoint node sets $P_k$ and $P_{k'}$ as $W(P_k,P_{k'})=\sum_{m\in P_k, m'\in P_{k'}} A_{mm'}$. Here, ${vol(P_k)}$ measures the volume of connections between $P_k$ and the rest of the graph and $W(P_k,P_{k'})$ measures the volume of connections between $P_k$ and $P_{k'}$. Notice that the rule we use to weight the edge encourages connections between very important and less important filters. We would like to partition the filters so that the knowledge of the teacher model can be distributed uniformly across the students. To this end, this work applies normalized cut, a prevalent spectral clustering method, to split the graph with the minimized weights of the cuts while encouraging the weights within each sub-graph to be large. In this way, it avoids the isolated nodes, i.e., filters of the final convolution layer, from the rest of the graph. Specifically, for a given number of partitions $K$, the normalized cut, denoted by $Ncut$, of $K$ partitions $\mathcal{P}=\{P_1,P_2,...,P_K\}$ is given by
\begin{equation}
    Ncut(P_1,P_2,...,P_k) = \frac{1}{2}\sum\limits_{k=1}\limits^K \frac{W(P_k,\overline{P}_k)}{vol(P_k)},
\end{equation}
where $\overline{P}_k$ represents the complementary set of $P_k$, i.e.,$\overline{P}_k=F-P_k$. The minimum of $\sum_{k}1/vol(P_k)$ is achieved if all ${vol(P_i)}$ coincide thus ensures the partitions are as balanced as possible. To facilitate the solution, we relax the Ncut minimization problem by involving indicator vectors and discarding the discreteness condition, which fits in with efficient spectral algorithms achieved by finding the smallest nonzero eigenvalue of the graph Laplacian and thresholding the entries of the corresponding eigenvector \cite{belkin2001laplacian}. The relaxed problem is given by
\begin{equation}
\begin{aligned}
\min_{H\in\mathcal{R}^{M\times K}} &tr(H^TZ^{-\frac{1}{2}}LZ^{-\frac{1}{2}}H) \\ 
s.t.\quad &H^TH=I. 
\end{aligned}
\end{equation}

Here, the Laplacian matrix, denoted by $L$, of the graph $\mathcal{F}$ is a symmetrical matrix and is defined as $L=Z-A$, where $Z$ is the degree matrix of vertices with element $z_i$ and $A$ is the adjacency matrix. The problem above is a standard trace minimization problem, and thus the filter partition problem is converted to optimize an $M$-by-$K$ indicator matrix $H$ whose element $h_{mk}>0$ if the filter $f_m$ belongs to partition $P_k$ and $h_{mk}=0$ otherwise. Basically, the solution is given by a matrix with the eigenvector associated with the $K$ smallest eigenvalue of normalized Laplacian, i.e., $L_{sym}=Z^{-1/2}LZ^{-1/2}$, as the columns, which can be obtained via eigenvalue decomposition of $L_{sym}$.


3) \textit{Student assignment}: Recall that a certain knowledge partition of the teacher can be learned by a proper student model that is replicated across the edge devices in each device group, such that the devices can generate output replicas of the students against any local failures. Based on the device grouping and knowledge partition strategy obtained before, assigning students across the edge devices calls for a three-dimensional matching among device groups, knowledge partitions, and student models. 

We would like to form a device-knowledge-student assignment in a manner that maximizes the inference accuracy while minimizing both computation time ($R_j/c_n^{core}$) and communication time ($Q_j/r_n$). Basically, a student model with a relatively complex structure and more parameters is more powerful to learn and mimic a large-sized knowledge partition, while the selected models are forced to meet specific memory- and FLOP-constraints of edge devices. Let $C^{para}(P_{k})$ indicate the size of the knowledge partition $P_{k}$. The accuracy performance can be further interpreted by the ratio between the size of any student model $S_j, \forall j$ and the size of any knowledge partition $P_k, \forall k$, denoted as $R_j/C^{para}(P_{k})$. It reflects the efficacy of using appropriately sized student models for learning specific knowledge, as employing a larger model for a smaller knowledge partition often yields better representation performance. 

Notice that, for a fixed device-knowledge pair, we can find the most appropriate student model from the student set $S$ by optimizing the accuracy-latency trade-off under the device's hardware constraints. Thus, the three-dimensional matching problem can be reduced to a bi-partite matching between $K$ device groups and $K$ knowledge partitions. We first narrow the set of applicable student models for each device group according to the memory constraints of the devices, where the narrowed set is denoted by $S_k$ with $S_k \subset S$. Accordingly, the device group $G_k$ may have one possible link with the $k'$-th knowledge partition, where the edge weight is defined as:
\begin{equation}\label{weight}
w(G_k,P_{k'})=\max \limits_{s_j \in S_k} \frac{R_j}{C^{para}(P_{k'}) (\frac{R_j}{c_n^{core}} + \frac{Q_j}{r_n})}, 
\end{equation}
wherein the weight is determined by the maximum accuracy-to-delay performance achievable across all feasible student model structures for device group $G_k$.

After constructing the weighted bipartite graph, the well-known Kuhn-Munkres algorithm can be used to give the optimal one-to-one pairing between device groups and knowledge partitions to maximize the sum weight. The detailed matching process is summarized in Algorithm~\ref{alg2}.

4) Complexity analysis: Algorithm 1 consists of three main functional modules: device grouping, knowledge partition, and student assignment, each contributing to the overall complexity of the algorithm. Device grouping employs follow-the-leader clustering procedures, typically scaling as $O(NK)$. The knowledge partition process involves the Normalized Cut algorithm with complexity $O(M^2)$. To optimize the student assignment strategy efficiently, we reduce the three-dimensional matching among devices, filters, and students into a bi-partite matching, optimally solvable using the Kuhn-Munkres algorithm with complexity $O(K^3)$. It's worth noting that this complexity can potentially be further reduced by employing alternative low-complexity bi-partite matching algorithms, albeit with varying extent of performance sacrifices. Consequently, the overall complexity of Algorithm 1 is $O(\max(NK,M^2,K^3))$.

\begin{algorithm}[!t]
\caption{Knowledge Assignment Algorithm} \label{alg2}
\hspace*{0.02in} {\bf Input:} Capacity similarity threshold $d^{th}$; Transmission failure probability threshold $p^{th}$\\
\hspace*{0.02in} {\bf Output:} Device groups $\mathcal{G}=\{G_1,...,G_k...,G_K\}$; Filter partitions $\mathcal{P}=\{P_1,...,P_k...,P_K\}$; Assignment strategy $\alpha_{kj},\forall k,j$\\
\hspace*{0.02in} {\bf Initialization:} $G_1 \leftarrow d_1$, $\mathcal{G} \leftarrow G_1$, $K=1$, $\mathcal{P} \leftarrow \emptyset$
\begin{algorithmic}[1]
\State {\bf Device grouping:}
\State Compute the centroid of every group $G_k$ as $\overline{G}_k, \forall k=1,...,K$ 
\For{$n = 2,3,...,N$}  
    \State Compute $sim(\overline{G}_k, d_n), \forall k=1,...,K$ 
    \For{$k = 1 \to K$}
        \If{$sim(\overline{G}_k, d_n)\leq d^{th}$ and $ \prod_{n:d_n\in G_k} (1-p_n)\leq p^{th} $}
            \State $G_k \leftarrow G_k \cup \{d_i\}$, update $\overline{G}_k$
            \State {\bf break} 
        \EndIf
    \EndFor
    \If{$d_i$ is unassigned}
        \State$G_{K+1} \leftarrow \{d_i\}$, update $\overline{G}_{K+1}$
        \State $\mathcal{G} \leftarrow \mathcal{G} \cup G_{K+1}$, $K \leftarrow K+1$
    \EndIf
\EndFor
\State {\bf Knowledge partition:}
\State Construct $\mathcal{F}=(F,E,A)$ with $A_{ij} = F_{ji} = a_i a_j \vert a_i - a_j \vert$ as the weight of the edge $e_{ij}$
\State Compute the degree matrix of $\mathcal{F}$ as $Z$
\State Compute the normallized Laplacian $L_{sym} \leftarrow Z^{-1/2}LZ^{-1/2}$ where $L=Z-A$
\State Compute the smallest $K$ eigenvectors $u_1,...,u_K$ of $L_{sym}$
\State Compute the indicator matrix $H \leftarrow [u_1,...,u_K], H \in \mathbb{R}^{M \times K}$
\State Cluster the rows in $H$ into $K$ clusters using K-Means and generate the corresponding knowledge partition strategy $\mathcal{P}=\{P_1,...,P_K\}$
\State {\bf Student assignment:}
\State Construct $S_k$ for every device group $G_k$ 
\For{$k,k' = 1,...,K$}
    \State Compute $w(G_k,P_{k'})$ through (\ref{weight})  
\EndFor
\State Use KM algorithm to obtain student assignment matrix $\Lambda$
\For{$k = 1,...,K$}
    \State Pick $s_j$ from $S_k$ with the maximum value of $\frac{C^{flop}_j}{C^{para}(P_{k'}) (\frac{C_j^{flops}}{c_n^{core}} + \frac{Q_j}{r_n})}$ and obtain the student model assignment strategy $\alpha_{jk}$ for the device group $G_k$
\EndFor
\State \Return $\mathcal{G}, \mathcal{P}, \mathcal{S}, \mathbb{\alpha}$
\end{algorithmic}
\end{algorithm}


\section{Performance Evaluation}
In this section, we evaluate the performance of the proposed cooperative inference mechanism RoCoIn via extensive simulations. Particularly, we compare RoCoIn with several baselines in terms of model complexity, failure resiliency and inference latency in various system conditions. We also explore the impact of cumulative transmission failure probability threshold and average transmission success probability on inference latency.

\subsection{Evaluation Setup}
We use the CIFAR-10 and CIFAR-100 datasets in our experiments, which are two of the most widely used datasets for machine learning research. The CIFAR-10 dataset contains 60,000 32x32 color images in 10 different classes, while CIFAR-100 dataset consists of 100 classes with 20 superclasses and makes the image classification task more complex to learn than that of CIFAR-10. We use WideResNet-16-4 and WideResNet-28-10 as the teacher networks, which are trained on both the CIFAR-10 dataset and CIFAR-100 dataset for image classification applications. For student networks, there are $2$ types of backbone models available to select, i.e. MobileNet and WideResNet. Mobilenet is a typical lightweight deep neural network specially designed for edge devices, and WideResNet is a variant of ResNet with decreased depth and increased width, which is far superior to their commonly used thin and very deep counterparts. We set $S=\{$WideResNet-22-1, WideResNet-16-1, MobileNet-v2$\}$ for CIFAR-10 and $S=\{$ WideResNet-16-3, WideResNet-16-2, WideResNet-22-1 $\}$ for CIFAR-100, where MobileNet-v2 and WideResNet-16-3 have the minimum and maximum computational loads and memory footprint, respectively. 
For all simulations, the number of edge devices that cooperatively perform DNN inference tasks is set to be $8$. We assume that the computational capacity of the edge device ranges from $5M$ to $30M$ FLOPS, and randomly set the transmitting rate $r_n$ in the range $[0.5,1]$ kbps for each device. Here, the student models are trained using the following loss function:
\begin{equation}\label{KDloss}
\begin{aligned}
    Loss(\theta_S)= & \begin{matrix}\underbrace{(1-\alpha)\mathbb{H}(y,P_S)+\alpha\mathbb{H}(P_T^\tau,P_S^\tau)}\\\text{KD loss}\end{matrix} \\ 
    & +\begin{matrix}\underbrace{\beta \sum \limits_{P_k \in \mathcal{P}} \left \| \frac{v_T^F(p)}{||v_T^F(p)||}-\frac{v_S^F(p)}{||v_S^F(p)||} \right\|_2^2}\\\text{AT loss}\end{matrix},
\end{aligned}
\end{equation}
where the first term is the standard knowledge distillation loss integrating hard\&soft-label cross-entropy loss, and the last term is the activation-transfer loss that reflects the error between activations of the teacher’s filters that belong to the given partition and activations of filters in the corresponding student.

In the runtime stage, we launch for the IoT edge cluster an image classification task on one image from CIFAR-10 or CIFAR-100, and take the average inference latency and accuracy of 100 repeated trials as the results. Particularly, we compare the performance of our RoCoIn against the following baselines, i.e.: 
\begin{enumerate}
    \item \textit{RoCoIn-G} employs a similar cooperative inference workflow as RoCoIn, but adopts a simple heuristic method to decide the knowledge assignment. 
    \item \textit{NoNN} partitions the knowledge of model equally and generates student models with the same architecture via knowledge distillation~\cite{bhardwaj2019NoNN}.
    \item \textit{HetNoNN} improves \textit{NoNN} by distributing teacher's knowledge based on devices’ memory and computing capabilities, but overlooking device grouping for resisting communication failures.
    \item \textit{Teacher} gives the performance of the original large model, which preserves the highest accuracy without knowledge loss but cannot be locally deployed on edge devices. 
\end{enumerate}

\subsection{Evaluation Results}

\begin{table}[!htbp]
  \begin{center}
    \caption{Results of Image Classification on CIFAR-10 Dataset.}
    \begin{tabular}{|l|l|c|c|c|}
    \hline
    \multicolumn{1}{|c|}{Method} & \multicolumn{1}{c|}{Model} & \begin{tabular}[c]{@{}c@{}}Parameters\\ (Largest)\end{tabular} & \begin{tabular}[c]{@{}c@{}}FLOPs\\ (Largest)\end{tabular}   & Accuracy \\ \hline
    Teacher & WideResNet16-4 & 2.75M & 507.84M & 91.86\%  \\ \hline
    \textbf{RoCoIn} & \textbf{WideResNet22-1} & \textbf{0.28M} & \textbf{48.58M} & \textbf{91.62\%}  \\ \hline
    RoCoIn-G & WideResNet22-1 & 0.28M & 48.58M  & 91.40\%  \\ \hline
    HetNoNN & WideResNet22-1 & 0.28M & 48.58M & 91.51\%  \\ \hline
    NoNN & WideResNet16-1 & 0.18M & 34.25M & 91.32\%  \\ \hline
    \end{tabular}
    \label{modelCifar10}
  \end{center}
\end{table}

\begin{table}[!htbp]
  \begin{center}
    \caption{Results of Image Classification on CIFAR-100 Dataset.}
    \begin{tabular}{|l|l|c|c|c|}
    \hline
    \multicolumn{1}{|c|}{Method} & \multicolumn{1}{c|}{Model} & \begin{tabular}[c]{@{}c@{}}Parameters\\ (Largest)\end{tabular} & \begin{tabular}[c]{@{}c@{}}FLOPs\\ (Largest)\end{tabular}   & Accuracy \\ \hline
    Teacher & WideResNet28-10 & 36.5M & 10.9G & 74.66\%  \\ \hline
    \textbf{RoCoIn} & \textbf{WideResNet16-3} & \textbf{1.56M} & \textbf{575.3M} & \textbf{72.42\%}  \\ \hline
    RoCoIn-G & WideResNet16-3 & 1.56M & 575.3M  & 72.31\%  \\ \hline
    HetNoNN & WideResNet16-3 & 1.56M & 575.3M & 71.68\%  \\ \hline
    NoNN & WideResNet16-2 & 0.71M & 260.1M  & 70.78\%  \\ \hline
    \end{tabular}
    \label{modelCifar100}
  \end{center}
\end{table}

\begin{figure}[t]
    \centering
    \subfigure[CIFAR-10]{ 
        \includegraphics[width=0.45\textwidth]{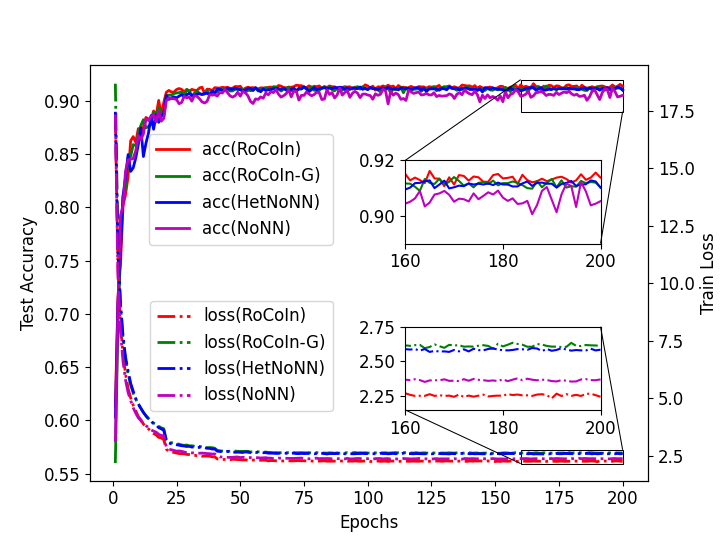}} 
    \hspace{1in} 
    \subfigure[CIFAR-100]{
        \includegraphics[width=0.45\textwidth]{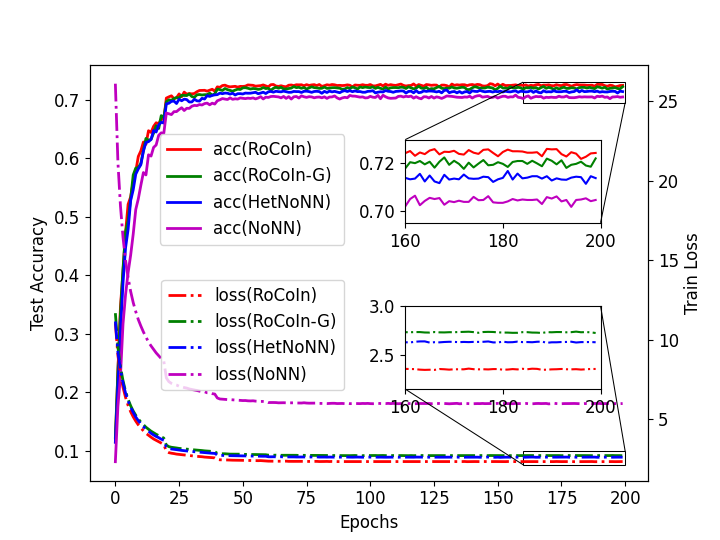}}
    \caption{Training performance.}
    \label{training}
\end{figure}

We first validate the superiority of distributed DNN inference in alleviating the IoT devices' computational load and evaluate the performance of the proposed RoCoIn scheme through a comparison with the other baselines. In Table \ref{modelCifar10} and \ref{modelCifar100}, we summarize the specifications for the locally-deployed DNN models with the different schemes for performing a certain inference task. As evidence, the distributed inference solutions result in student models with significantly fewer parameters and lower computational load compared to the original teacher model, allowing them to fit within the resource-constrained IoT devices. We notice that \textit{NoNN} may result in the smaller model sizes compared to RoCoIn, which is primarily due to the following factors: i) NoNN mandates that all edge devices deploy student models with identical structures, rendering it bottlenecked by low-end devices with low memory budgets. Thus, an extremely sparse student model will be selected for all the devices to cater to those ``stragglers'', even though the majority of devices can handle denser models with higher accuracy. ii) RoCoIn ensures the resilience of the cooperative inference system by strategically introducing redundancy when distributing the knowledge to the edge devices. With the fixed number of devices, the amount of ``knowledge'' that is required to be learned by individual devices increases, which may require more dense student models to attain satisfactory accuracy. Despite saving the memory and FLOPS, \textit{NoNN} considerably degrades the accuracy performance as it applies lightweight student models to all the devices subjecting to the tightest capacity constraint. It shows that RoCoIn can maintain high classification accuracy to the greatest extent among all the distributed solutions while ensuring lightweight computation and memory overhead. This is attributed to the appropriate knowledge assignment, which motivates those powerful devices to deploy parameter-rich student models that learn complex and important knowledge partitions. Fig. \ref{training} further reveals the training performance of the network of student models involved by different knowledge assignment schemes, where the accuracy and loss are calculated by aggregating the students' outputs and yielding the final predictions. The results validate that the test accuracy achieved by our RoCoIn is consistently higher than the baselines on both CIFAR-10 and CIFAR-100 datasets. 

\begin{figure}[t]
    \centering
    \subfigure[Inference latency]{ 
        \label{pth}
        \includegraphics[width=0.45\textwidth]{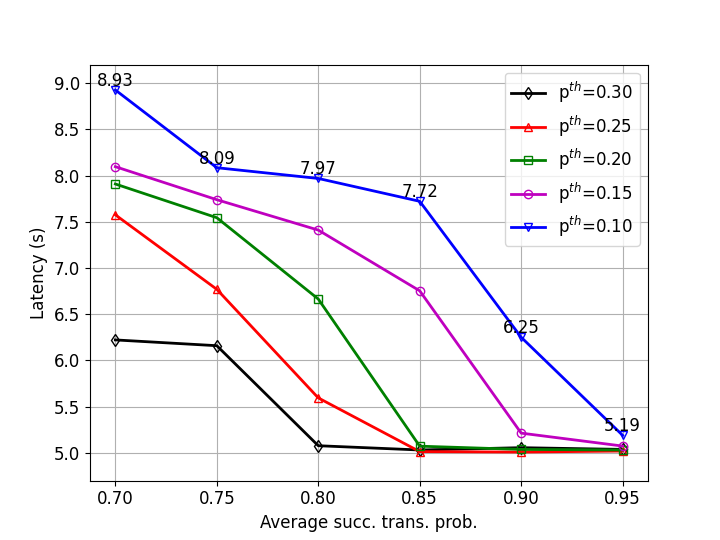}} 
    \hspace{1in} 
    \subfigure[Inference accuracy]{
        \label{pth_acc}
        \includegraphics[width=0.45\textwidth]{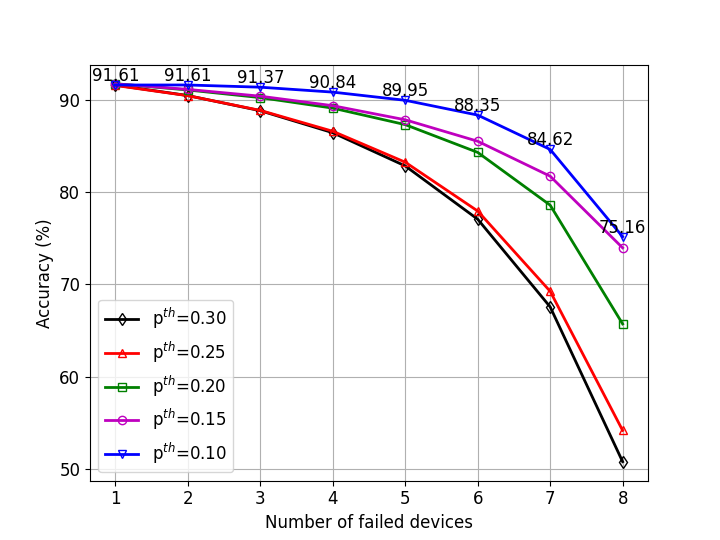}}
    \caption{RoCoIn's performance under different system configurations.}
\end{figure}

Basically, a high transmission requirement, i.e., a small transmission failure probability threshold $p^{th}$, for device grouping could probably increase the number of student replicas and thus results in highfailure resiliencee and low resource utilization. Fig.~\ref{pth} depicts the runtime inference latency under different configurations. We observe that the inference latency is non-increasing with the growing average success probability of the devices under different probability thresholds $p^{th}$. One reason for this is that, with a fixed threshold $p^{th}$, favorable communication conditions for devices can not only speed up the aggregation of local outputs but also potentially divide the teacher's knowledge into smaller partitions for more diverse distribution among devices, which reduces the need for redundant backup devices and ultimately decreases computational latency. A similar effect can also be achieved by increasing the threshold $p^{th}$, as verified in Fig.~\ref{pth}. As a consequence, the robustness of the RoCoIn system will be compromised as there aren't sufficient replicas in place to compensate for the loss of any student model's outputs. This conclusion can be drawn from Fig.~\ref{pth_acc}. 
In essence, an extremely small $p^{th}$ can make RoCoIn unable to find a feasible device grouping solution due to the strict target on the groups' cumulative transmission reliability. Yet the value of $p^{th}$ can be well designated to strike a balance between robustness and latency in practice. 

\begin{figure}[t]
    \centering
    \includegraphics[width=0.45\textwidth]{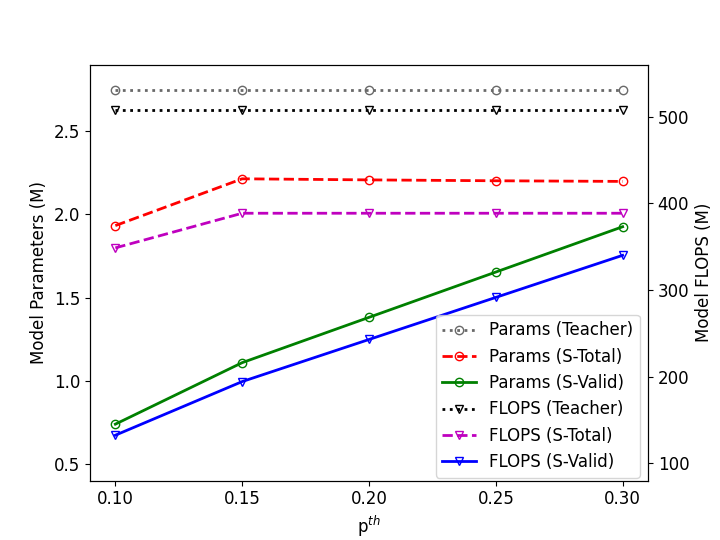}
    \caption{Student model profile for different redundancy mode.}
    \label{pth_resc}
\end{figure}

To further illustrate the impact of the $p^{th}$ on the robustness of RoCoIn system, we fix the average success probability of devices at 0.8 in this simulation and examine the inference accuracy of RocoIn with the existence of local failures under different thresholds $p^{th}$. Fig. \ref{pth_acc} gives the comparison result of different probability thresholds $p^{th}$, and the computational loads and parameters of corresponding student networks are concluded in Fig. \ref{pth_resc}. Here, we use S-Total and S-Valid to represent all student models including replicas and vital student models excluding replicas. A larger ratio of the valid value to the total value means better resource utilization efficiency. As can be observed in Fig. \ref{pth_acc} and Fig. \ref{pth_resc}, a smaller $p^{th}$ achieves better failure-resilience at the cost of lower resource utilization, which coincides with our basic design idea of RoCoIn. This result demonstrates that transmission failure probability threshold $p^{th}$ plays an important role in the performance of RoCoIn.

\begin{figure}[t]
    \centering
    \subfigure[CIFAR-10]{ 
        \includegraphics[width=0.45\textwidth]{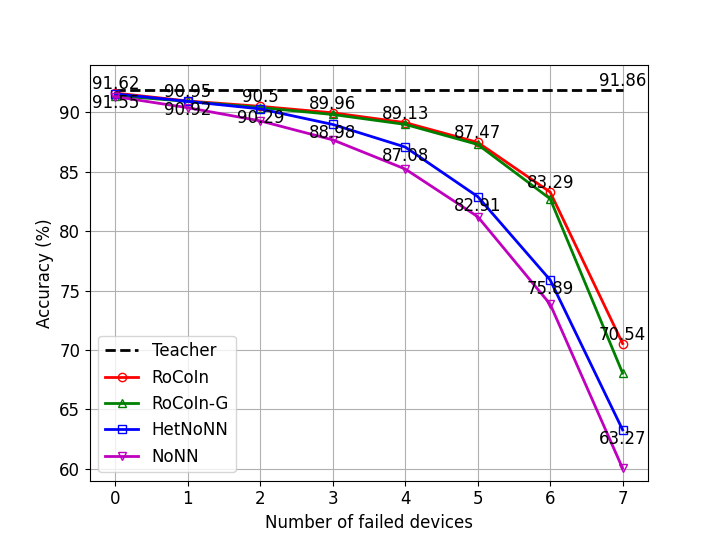}} 
    \hspace{1in} 
    \subfigure[CIFAR-100]{
        \includegraphics[width=0.45\textwidth]{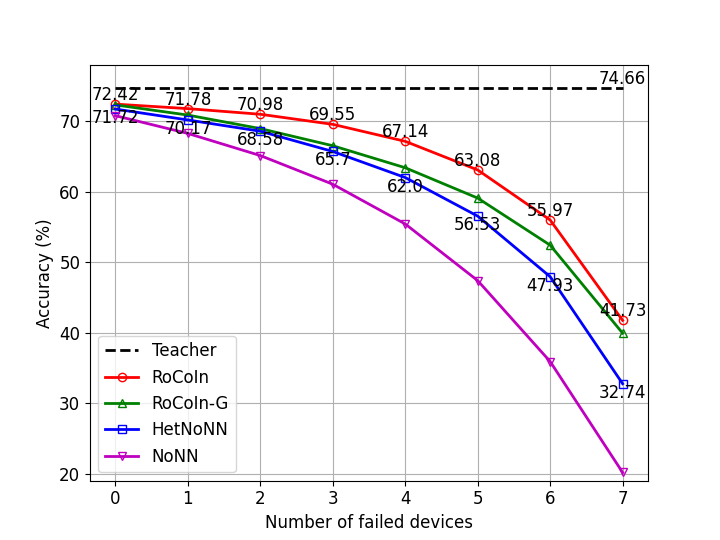}}
    \caption{Inference accuracy with failed devices.(w. known failure probability)}
    \label{w.prior.robust}
\end{figure}
\begin{figure}[t]
    \centering
    \subfigure[CIFAR-10]{ 
        \includegraphics[width=0.45\textwidth]{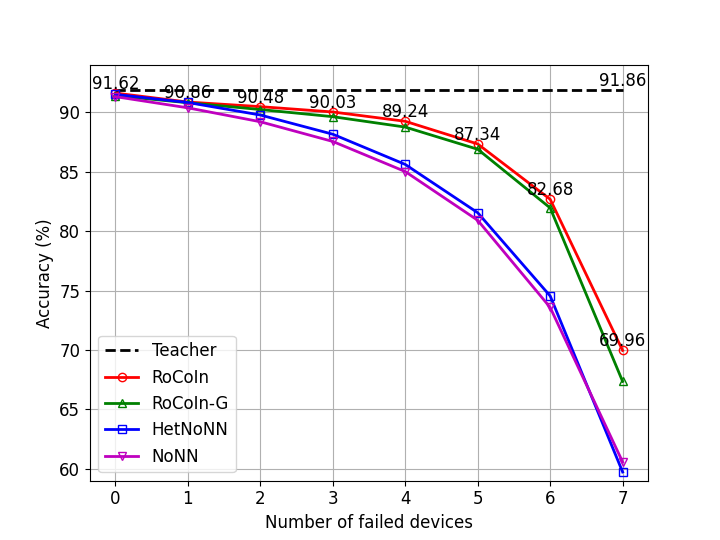}} 
    \hspace{1in} 
    \subfigure[CIFAR-100]{
        \includegraphics[width=0.45\textwidth]{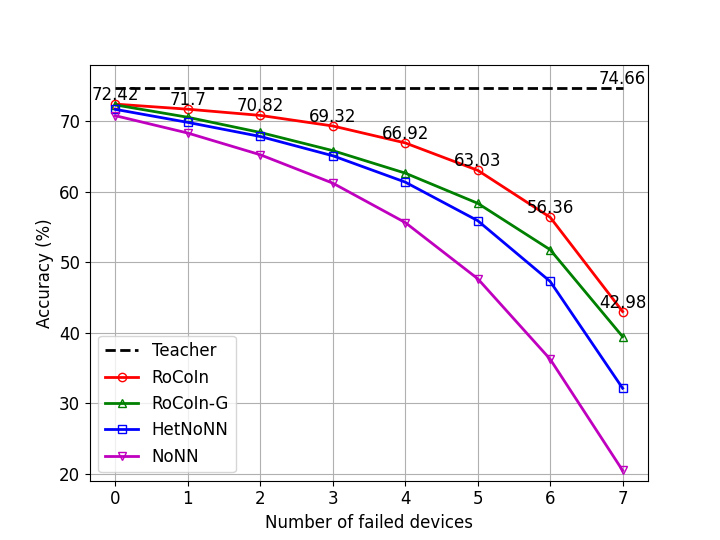}}
    \caption{Inference accuracy with failed devices. (w.o. known failure probability)}
    \label{robust}
\end{figure}

Fixing $p^{th}$ to $0.25$ and average success probability to $0.7$, we then examine the robustness of the distributed inference schemes against the cases that some devices are unavailable due to power depletion or communication failure, as shown in Fig.~\ref{w.prior.robust} and Fig.~\ref{robust}. Particularly, all schemes in Fig.~\ref{w.prior.robust} are configured with prior knowledge of device failure probabilities, whereas Fig.~\ref{robust} illustrates a more realistic scenario in which all schemes operate with unknown device failure probability distributions. Here, we emulate such local failures by simply zeroing out the inference results of the devices that were considered to experience failures when performing global aggregation. Then we study the impact of eliminating a different number of devices. For each setting, we randomly select from the device set a certain number of devices to delete and repeat for 30 trials, based on which we get the averaged inference accuracy to compare the failure-resilience performance of different distributed inference schemes. The results in Fig.~\ref{w.prior.robust} show that the absence of several devices degrades the cooperative inference accuracy with all the schemes, wherein our RoCoIn exhibit the most favorable performance in maintaining desirable accuracy. It is shown that, even if half of the devices fail to contribute to the inference outcome, RoCoIn can keep the classification accuracy over 88\% and 64\% for CIFAR-10 and CIFAR-100. That is to say, RoCoIn provides the failure-resilience guarantee when some of the local outputs get lost due to timeout or crash. In contrast, HetNoNN and NoNN are more sensitive to local failures, resulting in significant accuracy drops as the number of failed devices increases. We attribute this to the fact that RoCoIn tends to strategically group the devices for student replication and hence exhibits higher resilience. In the case where the prior knowledge of device failure probabilities is unknown, as shown in Fig.~\ref{robust}, RoCoIn exhibits a more significant performance gain compared to the baselines, due to its advantage of proactive replica deployment. This further underscores that in practical wireless distributed inference systems with environmental randomness, appropriate knowledge assignment with on-demand replication can effectively mitigate the detrimental impact of local errors on overall inference performance.

\begin{figure}[t]
    \centering
    \includegraphics[width=0.45\textwidth]{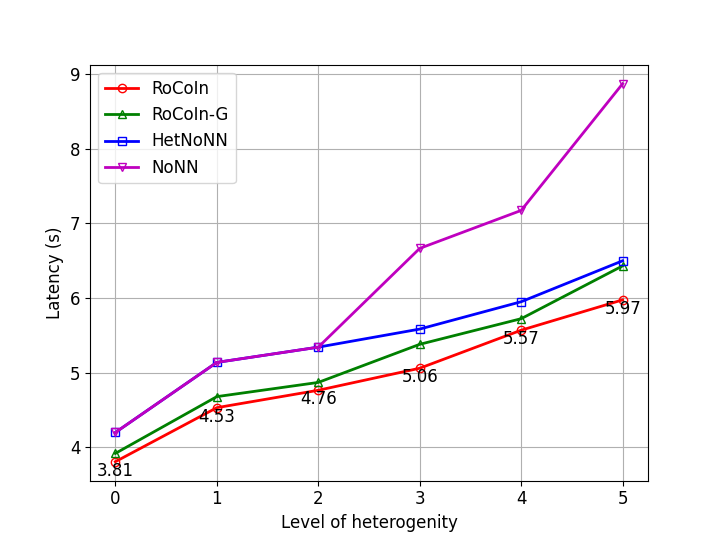}
    \caption{Inference latency under heterogeneous environments.}
    \label{hetero}
\end{figure}

\begin{table}[t]
  \begin{center}
    \caption{Level of Heterogeneity.}
    \begin{tabular}{|l|c|c|c|c|c|c|}
    \hline
    Heterogeneity level & 0 & 1 & 2  & 3 & 4 & 5 \\ \hline
    Range of FLOPS (M) & 0 & 10 & 15 & 20 & 25 & 30 \\ \hline
    Range of data rate (bps) & 0 & 100 & 200 & 300 & 400 & 500 \\ \hline
    \end{tabular}
    \label{heteroDef}
  \end{center}
\end{table}

Fig.~\ref{hetero} further evaluates the impact of the heterogeneity of devices' computational capacity and communication condition on the inference latency. We define a ``heterogeneity level'' to control the variation range of computing capability (FLOPS) and transmission rate among the devices. Here, we set six levels of heterogeneity, as described in Table~\ref{heteroDef}, randomly distribute the processing speed and transmitting rate of each device within the range. Fig.~\ref{hetero} elucidates that the high level of device heterogeneity has a negative impact on cooperative inference and impairs time efficiency. Among the schemes tested, \textit{NoNN} brings out the worst performance, especially for the cases of high heterogeneity, since it uniformly partitions and distributes the teacher's knowledge to the devices, ignoring their diverse capacity for handling workloads. In contrast, our proposed RoCoIn scheme, which integrates heterogeneity-aware knowledge assignment, outperforms the other baselines in overcoming the straggler issue in paralleled inference systems, regardless of the heterogeneity level. RoCoIn allows each device to run a well-selected student model that accommodates its computing and memory capacity, exhibiting greater adaptability than the others to cope with scenarios of high heterogeneity across devices while maintaining a low inference latency.

We also apply our RoCoIn scheme to an object detection task to assess its universality, utilizing the Yolov5 model~\cite{yolov5} and the VisDrone dataset~\cite{Visdrone}. This dataset comprises 288 video clips captured by various drone-mounted cameras, with manually annotated bounding boxes of targets such as pedestrians, cars, bicycles, and tricycles. To generate student models, we distill and parallelize the compute-intensive layers of the Yolo backbone and Neck modules to improve model compression efficiency. Here, \textit{Yolov5-BC} is a student architecture modified by the Yolov5 with a compressed backbone module, while \textit{Yolov5-BNC} compresses both backbone and neck modules. We evaluate the performance of RoCoIn with 2 devices and 3 devices, respectively, and present the results in Table~V. We observe that our RoCoIn can consistently reduces memory and computational costs to a certain extent due to computation parallelization, even with the more complex architecture of the Yolov5 model. While \textit{Yolov5-BC} requires keeping a relatively large student model at each device, it can achieve higher inference accuracy compared with \textit{Yolov5-BNC}. Although \textit{Yolov5-BC} necessitates maintaining a relatively large student model at each device, it achieves higher inference accuracy compared to \textit{Yolov5-BNC}. It can be envisioned that for complex DNN tasks with intricate model architectures, RoCoIn enables developers to determine which modules should be compressed and parallelized to strike a balance between accuracy and costs.


\begin{table}[!htbp]\label{res-detec}
  \begin{center}
    \caption{Results of Objective Detection on VisDrone2019 Dataset.}
    \begin{tabular}{|l|l|c|c|c|}
    \hline
    \multicolumn{1}{|c|}{Method} & \multicolumn{1}{c|}{Model} & \begin{tabular}[c]{@{}c@{}}Parameters\end{tabular} & \begin{tabular}[c]{@{}c@{}}FLOPs\end{tabular}   & mAP(0.5) \\ \hline
    Teacher & Yolov5-s & 7.23M & 16.6G & 48\%  \\ \hline
     \makecell{\textbf{RoCoIn} \\ (2 devices)} & \textbf{Yolov5-BC} &
    \begin{tabular}[c]{@{}c@{}}\textbf{4.97M} \\ \textbf{4.97M}\end{tabular}
     & \begin{tabular}[c]{@{}c@{}}\textbf{11.2G} \\ \textbf{11.2G}\end{tabular} & \textbf{41\%}  \\ \hline
     \makecell{\textbf{RoCoIn} \\ (2 devices)} & \textbf{Yolov5-BNC} &
    \begin{tabular}[c]{@{}c@{}}\textbf{1.76M} \\ \textbf{1.76M}\end{tabular}
     & \begin{tabular}[c]{@{}c@{}}\textbf{2.07G} \\ \textbf{2.07G}\end{tabular} & \textbf{28.2\%}  \\ \hline
    \makecell{\textbf{RoCoIn} \\ (3 devices)} & \textbf{Yolov5-BNC} & 
    \begin{tabular}[c]{@{}c@{}}\textbf{1.76M} \\ \textbf{0.98M} \\ \textbf{0.98M} \end{tabular} & \begin{tabular}[c]{@{}c@{}}\textbf{2.07G} \\ \textbf{1.27G} \\ \textbf{1.27G} \end{tabular} & \textbf{28.5\%} \\
    \hline
    \end{tabular}
    \label{modelCifar100}
  \end{center}
\end{table}

\section{Conclusion}
In this work, we have presented a RoCoIn scheme to enable failure-resilient distributed inference across multiple resource-constrained edge devices for offering deep neural network-based services. Considering the heterogeneous computing and communication capacity of devices, we have proposed to partition the knowledge of the original large model into independent modules and assign the computation workload of every knowledge module to edge devices with compressed student models, aiming to minimize the response latency of the distributed inference system. To make the cooperative inference system resilient to local failures, we use a clustering-based method to group the devices for redundant deploying the same student model and performing the corresponding computation workload. Extensive simulations have been conducted to evaluate RoCoIn's performance. The results have shown that the proposed mechanism exhibits great potential in accommodating the heterogeneity of edge devices and improving the system's robustness against local crash or timeout failures.


\bibliographystyle{IEEETran}
\bibliography{final.bib}

\end{document}